\documentclass[10pt,a4paper,final]{article}

\usepackage[utf8]{inputenc}

\usepackage{geometry}
\usepackage[backend=biber,style=numeric-comp,natbib=true,url=false,doi=true,sorting=none,maxnames=99]{biblatex}
\DeclareUnicodeCharacter{0301}{\'{e}}
\AtEveryBibitem{\clearfield{month}} 
\renewbibmacro{in:}{%
\ifentrytype{article}{}{\printtext{\bibstring{in}\intitlepunct}}} 
\DeclareFieldFormat{pages}{#1} 

\renewbibmacro*{journal+issuetitle}{%
  \usebibmacro{journal}%
  \setunit*{\addspace}%
  \iffieldundef{series}
    {}
    {\newunit
     \printfield{series}%
     \setunit{\addspace}}%
  \setunit{\addspace}%
  \usebibmacro{issue+date}%
  \setunit{\addcolon\space}%
  \usebibmacro{issue}%
  \setunit{\addcomma\space}%
  \usebibmacro{volume+number+eid}%
  \newunit}

\DeclareFieldFormat[article,periodical]{number}{\mkbibparens{#1}}
\renewbibmacro*{volume+number+eid}{%
  \printfield{volume}%
  \printfield{number}%
  \setunit{\addcomma\space}%
  \printfield{eid}}

\renewbibmacro*{issue+date}{%
  \iffieldundef{issue}
    {\usebibmacro{date}}
    {\printfield{issue}%
     \setunit*{\addspace}%
     \usebibmacro{date}}%
  \newunit}
\renewbibmacro*{journal+issuetitle}{%
  \usebibmacro{journal}%
  \setunit*{\addcomma\space}%
  \iffieldundef{series}
    {}
    {\newunit
     \printfield{series}%
     \setunit{\addspace}}%
  \usebibmacro{volume+number+eid}%
  \setunit{\addcomma\addspace}%
  \usebibmacro{issue+date}%
  \setunit{\addcolon\space}%
  \usebibmacro{issue}%
  \newunit}
\usepackage{graphicx}
\usepackage{float}
\usepackage{amsmath}
\usepackage{mathtools}
\usepackage{color,colortbl}
\usepackage[breaklinks=true,colorlinks=true,linkcolor=blue,urlcolor=blue,citecolor=blue]{hyperref} 
\usepackage{cleveref}
\usepackage[subpreambles=true]{standalone}
\usepackage[english]{babel}
\usepackage{import}
\usepackage{setspace}
\usepackage[useregional]{datetime2}
\usepackage{textcomp}
\usepackage{gensymb}
\usepackage{subcaption}
\usepackage{comment}
\usepackage{authblk}

\usepackage{chemformula}  
\usepackage{braket}
\usepackage{diagbox}
\usepackage{multirow}
\usepackage{tikz,xcolor,hyperref}
\hypersetup{
    breaklinks=true,
    colorlinks=true,
    linkcolor=blue,
    citecolor=blue,      
    urlcolor=blue,
}
\usepackage{csquotes} 
\usepackage{pifont}
\definecolor{amber(sae/ece)}{rgb}{1.0, 0.49, 0.0}
\definecolor{airforceblue}{rgb}{0.36, 0.54, 0.66}
\definecolor{alizarin}{rgb}{0.82, 0.1, 0.26}
\definecolor{antiquefuchsia}{rgb}{0.57, 0.36, 0.51}
\definecolor{applegreen}{rgb}{0.55, 0.71, 0.0}
\definecolor{blue(pigment)}{rgb}{0.2, 0.2, 0.6}
\definecolor{black}{rgb}{0.0, 0.0, 0.0}
\definecolor{BrightGray}{rgb}{.94,.94,.94}
\definecolor{maroon}{cmyk}{0,0.87,0.68,0.32}

\definecolor{lime}{HTML}{A6CE39}
\DeclareRobustCommand{\orcidicon}{%
	\begin{tikzpicture}
	\draw[lime, fill=lime] (0,0) 
	circle [radius=0.16] 
	node[white] {{\fontfamily{qag}\selectfont \tiny ID}};
	\draw[white, fill=white] (-0.0625,0.095) 
	circle [radius=0.007];
	\end{tikzpicture}
	\hspace{-2mm}
}

\foreach \x in {A, ..., Z}{%
	\expandafter\xdef\csname orcid\x\endcsname{\noexpand\href{https://orcid.org/\csname orcidauthor\x\endcsname}{\noexpand\orcidicon}}
}

\title{Controlling high-harmonic generation from strain engineered monolayer phosphorene}

\author[1,2]{Saibabu Madas\orcidB{}}
\author[1]{Subhendu Kahaly\orcidA{}\ding{41}}
\author[1]{Mousumi Upadhyay Kahaly\orcidF{}\ding{41}}
\affil[1]{ELI ALPS, ELI-HU Non-Profit Ltd., Wolfgang Sandner utca 3., Szeged 6728, Hungary}
\affil[2]{Institute of Physics, University of Szeged, D\'om t\'er 9, H-6720 Szeged, Hungary}

\affil[\ding{41}]{email: \textcolor{blue}{mousumi.upadhyaykahaly@eli-alps.hu}; \textcolor{blue}{subhendu.kahaly@eli-alps.hu}}

\date{}

\begin{document}
\maketitle

\section*{Abstract}
\begin{sloppypar}
Phosphorene, a well-studied 2D allotrope of phosphorus, features unique properties such as widely tunable bandgap, high carrier mobility, and remarkable intrinsic in-plane anisotropy. Utilizing these structural and electronic properties, we investigate ultrafast electron dynamics and high harmonic generation (HHG) from phosphorene subject to band structure engineering through external strain, based on \textit{ab initio} time-dependent density-functional theory approach. 
We show that strong field processes in such systems can be optimized and controlled by biaxial tensile and compressive strain engineering, that results in electronic structure modification.  While $-10\%$ strain resulted in closing of band gap, $2\%$ strain increased the gap by $22\%$ with respect to $0.9 ~eV$ in pristine phosphorene, consequently affecting  the high harmonic yield. 
With reduction of gap, by applying strain from $2\%$ to $-10\%$, the valence band near $\Gamma-$point becomes more flat and discreet, resulting in large electronic density of states and enhanced electronic excitation, which reflects in their ultrafast sub-cycle dynamics under laser excitation. 
Moreover, due to its intrinsic in-plane anisotropy, harmonic yield with laser polarization along the armchair (AC) direction is found to be higher than that of the zigzag (ZZ) direction for all the strain cases. Nearly, an order of magnitude enhancement of harmonic intensity is achieved  for $-10\%$ strain along AC direction.
The current study expands the research possibilities of phosphorene into a previously unexplored domain, indicating its potential for future utilization in extreme-ultraviolet and attosecond nanophotonics, and also for efficient table-top HHG sources.

\section{Introduction}

The advent of high-harmonic generation (HHG) from atomic system \cite{McPherson1987} had ignited massive interest in both experimental \cite{Corkum1993,Krause1992} and theoretical  \cite{Lewenstein1994} attosecond science research community. It created the possibility of extending the then existing techniques utilized in the attosecond science research field for atomic systems to condensed phase systems. HHG from solids is an emerging field in attosecond science and material physics \cite{Ghimire2010,Luu2015,You2016,Han2016}, demonstrating great prospects as compact attosecond light-source technologies \cite{Midorikawa2022, Nayak2019}, and also as novel ultrafast spectroscopy techniques capable to probe electronic structures \cite{Damascelli2004}. 
Different solid structure exhibit their own distinct band structure \cite{KahalyJPCC2008}, thereby presenting diverse prospects for interband and intraband dynamics \cite{Kaloni2012}, which have been identified as crucial factors at the fundamental level for solid-state HHG \cite{Goulielmakis2022}. However, the comprehensive understanding of the coupling mechanism between interband and intraband processes remains incomplete and difficult to be understood with conventional atomic three-step model often used to address HHG processes from atomic systems. 

 
The mechanism of HHG in condensed phase systems has main contribution from interband and intraband processes: 
(i) an electron-hole pair is generated due to the excitation of an electron from the valence band to the conduction band, usually by means of tunnelling near the band gap minimum, resulting in interband transition. (ii) The solid crystal momentum will follow the time-dependent vector potential (according to the acceleration theorem \cite{Bloch1929,Houston1940}) resulting in carrier motion in the bands, which will lead to nonperturbative intraband harmonics emission \cite{Ghimire2010}. (iii) Simultaneously, interband harmonic radiation can occur due to electron recombination, with frequencies corresponding to the instantaneous band gap \cite{Vampa2014}. The complicated relationship between intraband and interband processes can be controlled and utilized to enhance the HHG emission \cite{Schubert2014,Higuchi2014}. However, they are non-trivially coupled, with the interplay between them being either co-operative or competitive \cite{Vampa2014,Golde2008}; hence their contribution to the overall HHG yield should be investigated. In this aspect, quantum confinement in solid-state systems can be used as an additional useful means to control interband and intraband dynamics \cite{Nakagawa2022}, in parallel to optical tuning methods, such as tuning laser field polarization direction \cite{Kaneshima2018} and ellipticity \cite{Tamaya2016a}. 


Quantum confined systems \cite{UpadhyayKahaly2017, Kahaly2008}, for instance, two dimensional (2D) material's optical and electronic properties are very sensitive to the external perturbations, because of their atomic thickness \cite{TancogneDejean2018}. 
Amongst wide range of 2D materials synthesized in recent past \cite{MendozaSnchez2016}, monolayer black phosphorene (BP) displays band gap between that of graphene ($0 ~eV$) \cite{Novoselov2004} and transition-metal dichalcogenides (TMDs) ($1~ eV-2 ~eV$) \cite{Wang2012}, resulting in efficient photo-response in the infrared (IR) range. 
Hence, due to its highly anisotropic characteristics \cite{Madas2019}, together with its high carrier mobility, significant band gap, and excellent mechanical properties that can withstand significant amount of deformation before failure, monolayer black phosphorene is emerging as a viable contender in the competitive field of 2D materials, in particular for optoelectronics. Regardless of the fact that phosphorene is highly unstable in ambient temperature, with several emerging ambient stabilization techniques for phosphorene \cite{Zeng2021}, this system offers a new test bed for exploring its potential for IR optoelectronics and high harmonic spectroscopy. 
One interesting feature of this 2D material is the possibility to tune the band gap by structural engineering \cite{UpadhyayKahaly2017, KahalyJPCC2008}. The effects of different types of strains (uni-axial or biaxial) along the two anisotropic lattice vectors on the ground state properties of phosphorene using density functional theory (DFT) \cite{Kohn1965} have been extensively studied so far \cite{Sa2014,Seo2017,Yarmohammadi2020}. Experimentally, biaxial tensile strain can be applied through differential thermal expansion method that relies on a large difference in thermal expansion coefficients between the 2D material and the substrate. When a 2D material strongly bonds to the substrate and the system is heated or cooled, the expansion difference of these materials result in the creation of a uniform tensile biaxial strain \cite{Ahn2017}. This thermal strain will be evenly distributed in two perpendicular directions if the substrate is either amorphous, cubic, or possesses a (0001) surface in a hexagonal crystal, or a (001) surface in a tetragonal crystal. In addition, piezoelectric materials under the influence of an external electric field can be used for applying biaxial compressive and tensile strains in 2D materials  \cite{Hui2013,Ding2010,Jie2013}. In this method, a 2D material is placed on a hybrid substrate that includes a PMN-PT (lead magnesium niobate-lead titanate) layer, which experiences thickness variations when subjected to an external electric field. With the application of an appropriate electrical bias, the substrate extends vertically and contracts horizontally. This, in turn, induces a uniform biaxial compressive strain onto the 2D materials. On the other hand, when the bias direction is reversed, it can induce tensile strain as well. Phosphorene can withstand tensile strains up to 30\% and 27\% along the armchair (AC) and zigzag (ZZ) direction, respectively \cite{Wei2014}. Hence, strain engineering can be a practicable way to tune the electronic properties of phosphorene. Can the utilization of strain be a potent means to manipulate ultrafast electron dynamics and consequent optical properties in such materials?   
 
 In this study, using a full \textit{ab initio} approach based on DFT \cite{Kohn1965} and real-time time-dependent density functional theory (TDDFT) \cite{Marques2004}, we demonstrate the strong-field nonlinear response and the pronounced sensitivity of the HHG process to the modulation of band structures in monolayer black phosphorene, that can be achieved through strain engineering. We find that substantial enhancement of harmonic intensity by up to an order of magnitude is possible for some compressive strained cases and explore the possible origins behind the response/phenomena. We investigate the effects of intrinsic anisotropy of phosphorene by applying electric field along the AC and ZZ directions on the HHG and total harmonic yield. Our results indicating the tunability of HHG and total yield through both strain engineering and in-plane anisotropy can be useful for exploring phosphorene towards extreme-ultraviolet and attosecond nanophotonics. 


 
 This article is organized in the following way. The simulation approaches and parametric details are described in \cref{methods}. Then, we present our findings for the pristine and strained phosphorene systems, their laser-induced response and HHG properties in \cref{results}. In \cref{conclusions} conclusions are summarized.

\begin{figure}[t] 
\centering
\includegraphics[width=0.72\columnwidth]{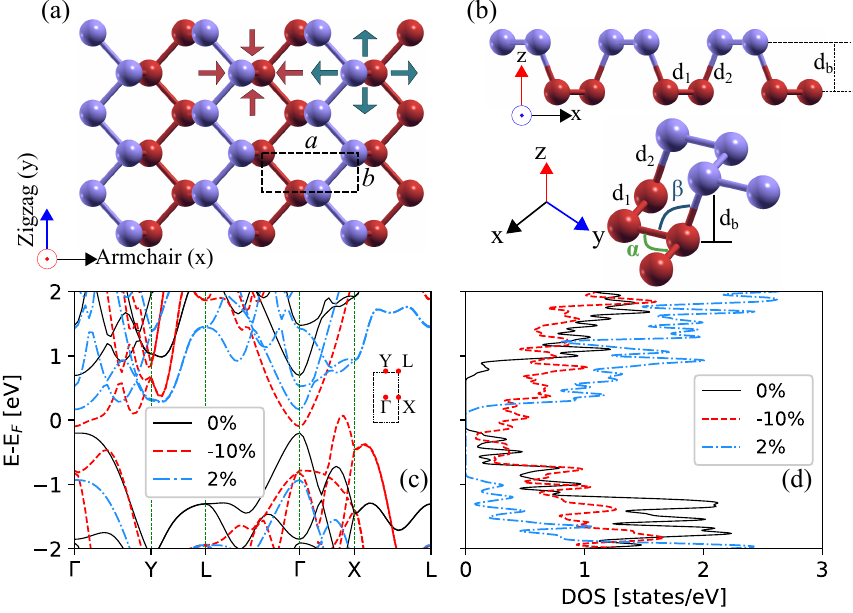}
\caption{Atomic structure of monolayer black phosphorene with it's crystallographic directions x (Armchair), y (zigzag). Purple and red colored spheres represent phosphorous atoms located in different planes within a single puckered layer. (a) Top view with representative unit cell structure (black dashed rectangle) and biaxial compressive (- ve) and tensile (+ ve) strains denoted by red and dark green arrows, respectively. (b) Side view with geometrical parameters: $d_1$ and $d_2$ are the bond lengths, $d_b$ is the buckling length. $\alpha$ and $\beta$ are the bond angles. (c) Electronic band structure along high symmetry directions (green vertical dashed lines), and (d) corresponding DOS of relaxed pristine (black solid curve), $-10\%$ compressive (red dashed curve), $2\%$ tensile (blue curve) strained phosphorene. Fermi level ($E_F$) is shifted to zero in (c) and (d). Inset in (c) shows irreducible Brillouin zone (BZ) with high symmetry points of phosphorene.}
\label{fig1}
\end{figure}

\section{Method}\label{methods}

In our work, structural optimization and electronic structure calculations of pristine and strained phosphorene systems were carried out within DFT \cite{Kohn1965} using a plane-wave basis set and ultrasoft pseudo-potential, as implemented in quantum espresso (QE) \cite{Giannozzi2009} package. The generalized gradient approximation (GGA) is used  for the exchange and correlation energy as proposed by Perdew, Burke, and Ernzerhof (PBE) \cite{Perdew1996}. To simulate a monolayer of black phosphorous, a rectangular unit cell with periodic boundary conditions is created. To avoid the interaction between system and the periodic images of slabs along the \textit{z}-direction, fairly large vacuum of 16 $\text{\AA}$ is created. The wave functions were expanded in plane waves up to an energy cutoff of $60~ eV$. For Brillouin zone (BZ) integration, we used the \textbf{k}-point sets generated by the $12\times12\times1$ Monkhorst-Pack \cite{Monkhorst1976} mesh. All atoms are relaxed till the final force exerted on each atom is less than $0.001~eV/\text{\AA}$ and also until the electronic energy convergence of $10^{-6}~eV$ is reached.

At room temperature, the most stable phosphorus allotrope is the black phosphorus, which was synthesized by Bridgman in 1914  \cite{Bridgman1914}. Black phosphorus is a layered material in which each atomic layers are stacked together by the vdW interactions. Unlike graphene, which is a planar layer of carbon atoms, black phosphorene (monolayer of black phosphorus) is a non-planar layer of phosphorous atoms. Four phosphorous atoms are present in the unit cell, among them, two atoms lie in one plane and the rest lie in another parallel plane, which results in a puckered honeycomb structure with each phosphorous atom covalently bonding with three adjacent atoms, as a result it has an anisotropic crystal structure, as shown in the Figure \ref{fig1}(a,b). In our calculations, we have optimized the lattice constants of black phosphorene, $a=4.62~\text{\AA}$ and $b=3.29~\text{\AA}$, which are well in agreement with the experiment \cite{Cartz1979} and the existing literature \cite{Qiao2014,Yasaei2015}. The relaxed geometrical parameters of phosphorene (shown in Figure \ref{fig1}b), such as the in-plane ($d_1$),  out-of-plane ($d_2$) bond lengths, inter-planar distance or the buckling length ($d_b$), angle among in-plane atoms ($\alpha$), and the angle among out-of-plane atoms ($\beta$) are tabulated in Table \ref{geometry_parameters}. 

The optimized pristine and strained phosphorene systems from QE calculations are then used in Octopus package \cite{Andrade2015,Andrade2012} for further time-dependent calculations, after ensuring force and energy convergence criteria are met. In order to investigate the laser-driven time dependent electron dynamics in pristine and the strained phosphorene systems, we employed semi-periodic boundary conditions, with a simulation box size of 30 Bohr along the non-periodic direction (along \textit{z} axis), which includes 3 Bohr of absorbing regions on both the sides of the phosphorene. This is a precautionary measure to prevent unphysical reflection of field-accelerated electrons at the border of the simulation box. The absorbing regions are treated using the complex absorbing potential (CAP) method \cite{DeGiovannini2015}, with a CAP height of -1 atomic units (a.u.). The real-space cell was sampled along all the three directions by a uniform grid spacing of 0.46 Bohr and $38\times46$ Monkhorst-Pack \cite{Monkhorst1976} \textbf{k}-points grid was used to sample the 2D Brillouin zone. To study the nonlinear processes, the time evolution of the wave functions and the estimation of time-dependent electronic current were calculated by propagating the Kohn-Sham  equations in real time and real space within TDDFT implemented within Octopus package \cite{Andrade2015,Andrade2012}, within GGA PBE exchange and correlations \cite{Perdew1996}. 

\section{Results and Discussion}\label{results}
\subsection{Crystal structure} 
In this work, we applied two types of biaxial strain: compressive (- ve) and tensile (+ ve) by tuning the in-plane lattice constants \textit{a} and \textit{b} of pristine phosphorene along both \textit{x} or armchair (AC) and \textit{y} or zigzag (ZZ) directions, as shown in Figure \ref{fig1}(a). Strain ($\varepsilon$) is defined as $\varepsilon = (a_\varepsilon-a_0)/a_0$, where $a_0$ and $a_\varepsilon$ are the pristine and strained phosphorene lattice constants, respectively. As it is evident from Table \ref{geometry_parameters}, applied strains affect the geometrical parameters (see Figure \ref{fig1}b): $d_1$, $d_2$, $d_b$, $\alpha$ and $\beta$ are slightly changed due to the application of tensile strain ($\varepsilon=2\%$), where the lattice constants \textit{a} and \textit{b} increases due to the elongation. On the other hand, for $\varepsilon=-10\%$ strain, the lattice constants \textit{a} and \textit{b} decreases resulting in significant change of the geometrical parameters. Note that such strains are well below the damage threshold as reported in the existing literature \cite{Wei2014}. 
\begin{table}[ht!]
\small
\newcolumntype{C}{>{\centering\arraybackslash}c}
\centering
\begin{adjustbox}{max width=1\columnwidth}{\begin{tabular}{|c|c|c|c|c|c|c|c|c|}\hline
\rowcolor{BrightGray} {Strain} & \emph{a} ($\text{\AA}$) & \emph{b} ($\text{\AA}$)&\emph{d$_1$} ($\text{\AA}$)&\emph{d$_2$} ($\text{\AA}$)&\emph{d$_b$} ($\text{\AA}$)&$\alpha$ (deg)&$\beta$ (deg)&$E_g$ (\textit{eV})\tabularnewline \hline
\rowcolor{green!5}  0\%&4.62 &3.29&2.22&2.26&2.11&104.08&95.6&0.9  \tabularnewline \hline
\rowcolor{white}   -10\%&4.16  &2.96&2.13&2.28&2.21&97.07&88.15&- \tabularnewline \hline
\rowcolor{green!5}  2\%&4.72  &3.36&2.24&2.26&2.09&104.76&97&1.1 \tabularnewline \hline
 
\end{tabular}}
\end{adjustbox}
\caption{Relaxed geometrical parameters and band gap ($E_g$) values of pristine and strained phosphorene systems. Parameters such as lattice constants \emph{a}, \emph{b}, bond lengths $d_1$, $d_2$, buckling length $d_b$, bond angles $\alpha$, $\beta$ are depicted in Figure \ref{fig1}(b).}
\label{geometry_parameters}
\end{table}

As tabulated in Table \ref{geometry_parameters}, $-10\%$ strain induces an increase in $d_2$ and $d_b$ by 0.8\% and 4.7\%, respectively, and decrease in $d_1$, $\alpha$ and $\beta$ by 4\%, 6.7\% and 7.7\%, respectively, compared to the pristine case. 

\subsection{Strain induced electronic properties}

Now we explore the differences in electronic band structure and highlight the important changes in the electronic properties as a function of the applied biaxial strain. For relaxed black phosphorene structure, we obtain a direct band gap of $0.9 ~eV$ at $\Gamma-$point, while the reported band gap from experiments is between $2.05-2.2 ~eV$ \cite{Liang2014}. As shown in Figure \ref{fig1}(c) (in black) for pristine phosphorene, the maximum contribution to the valence band maximum (VBM) and conduction band minimum (CBM) near the $\Gamma-$point is from the $p_z$ orbital, as revealed through projected bands and partial densities of states (PDOS) analysis [Figure \ref{Si1}(a,d)]. For $\varepsilon=2\%$, the bandgap increases to $1.1 ~eV$, with slight downshift of both valence bands (VB) and conduction bands (CB). Both the VBM and CBM originating still from $p_z$ orbital at $\Gamma-$point [see Figure \ref{Si1}(c,f)]. However, for $\varepsilon=-10\%$, the compressive strain prominently shortens the in-plane interatomic P-P bond length ($d_1$) and increases the in-plane $p_{x,y}$ orbital contributions. Consequently, superposition of the atomic orbitals leads to shift in the energy of the states and an indirect band gap closure due to the crossing of VBM and CBM at $\Gamma\longrightarrow X$ path as denoted with red dashed lines in Figure \ref{fig1}(c). As projected bands and PDOS analysis revealed that while CBM still mainly originates from the $p_z$ orbitals [see Figure \ref{Si1}(b,e)] at $\Gamma-$point, the in-plane orbitals $p_x~\text{and}~p_y$ contributions become dominant in the VBM along the $\Gamma\longrightarrow X$ path and crosses the Fermi level [\ref{Si1}(b,e)]. Consequently, more electronic density of states (DOS) near the VBM and CBM are obtained as compared to the other two cases, as shown in Figure \ref{fig1}(d). These effects result in compressive strain-induced semiconductor to  metal transition in phosphorene. The calculated band gap values of pristine and strained phosphorene systems (tabulated in Table \ref{geometry_parameters}) are well in agreement with the existing literature \cite{Wang2015}.

 
\subsection{Effect of strain on nonlinear response}
\begin{figure}
    \centering
    \includegraphics[width=.98\textwidth]{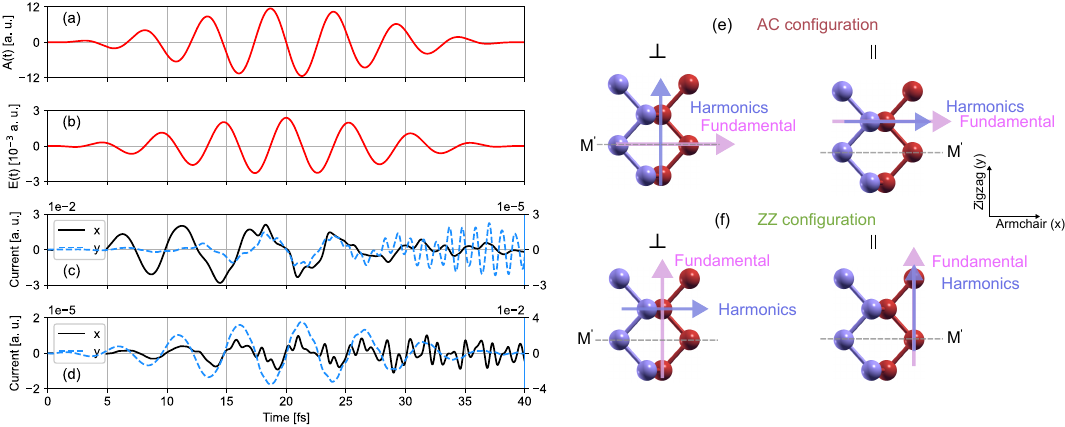}
    \caption{(a) The applied vector potential $\pmb{A}(t)$ and the corresponding (b) electric field $\pmb{E}(t)$ along with the induced electronic currents in pristine phosphorene that are parallel ($\parallel$) and perpendicular ($\perp$) to the incident laser polarization direction for the AC (c) and ZZ (d) configurations. The arrangement of (e) armchair (AC) and (f) zigzag (ZZ) configuration defined with respect to the laser polarization or fundamental field direction and the harmonic fields. The pink and blue arrows refers to the fundamental field and the harmonic fields, respectively. The grey dashed line represents the mirror plane (M$'$) of the phosphorene. Note that the current follows the vector potential; at peak electric field, the current is zero for both AC and ZZ in (c,d).}
    \label{fig2}
\end{figure}

Such strain-induced local structural modifications \cite{UpadhyayKahaly2011} and resulting electronic properties has direct impact on the time-dependent nonlinear responses in phosphorene. To probe this, we consider a driving laser field which is spatially uniform and is described in the velocity gauge, i.e. in terms of a vector potential [$\pmb{A}(t)$] to treat purely in-plane perturbation, given as
 
 
\begin{equation}\label{vec_pot}
     \pmb{A}(t) = -\frac{cE_0}{\omega_0}f(t)\cos{(\omega_0t+\phi)}~, 
\end{equation}

\noindent
where \textit{c} is the speed of light, $E{_0}$ is the electric field amplitude and $\omega_0$ is the carrier frequency of the driving laser pulse. The envelop of the vector potential $f(t)$ considered in this work is sine squared, that is defined as $f(t)=sin^2(\pi t/2\tau)$ for $0<t<2\tau$, or else $f(t)=0$. $\tau = 20~ fs$ is the pulse duration at full width half maximum (FWHM). $E_0$ in atomic units is defined in terms of peak intensity of the laser ($I_0$) and is given as $E_0 = \sqrt{I_0/3.51\times 10^{16}}$, where $I_0=2\times10^{11}~Wcm^{-2}$. All calculations are performed using a laser beam with a central wavelength ($\lambda$) of $1600~ nm$ (corresponding to a photon energy of $0.77~ eV$), and with a zero carrier envelope phase ($\phi=0$). The corresponding electric field is defined as $\pmb{E}(t) = -\partial \pmb{A}/(c~\partial t)$. The applied laser waveform, i.e. the vector potential and the corresponding electric field are depicted in Figure \ref{fig2}(a) and Figure \ref{fig2}(b), respectively. The pulse parameters are chosen to represent Mid-IR laser regime as available in ELI ALPS facility \cite{Khn2017,Charalambidis2017} and is suitable to probe HHG from solid-state materials. Note that we assume dipole approximation and the contributions from magnetic component of the electromagnetic field is ignored. Also any other relativistic terms, such as spin-orbit coupling are not considered within our adopted framework. Our systems are excited by the laser polarized either along AC or ZZ direction, where $\overline{\Gamma X}$ and $\overline{\Gamma Y}$ shown in the BZ (Figure \ref{fig1}c), corresponds to the AC and ZZ directions in real-space, respectively. 

In order to calculate the harmonic spectrum, we obtain total electronic current in the system, $\pmb{J}(t)=\partial/\partial t\int_{\Omega}d^3\pmb{r}\pmb{j}(\pmb{r},t)$, which is obtained by integrating the microscopic electronic current density [$\pmb{j}(\pmb{r},t)$] over the unit cell, where $\Omega$ is the unit cell volume \cite{TancogneDejean2017}. The microscopic electronic current $\pmb{j}(\pmb{r},t)$ can be written as 
\begin{equation}\label{electronic_current}
 \frac{\partial}{\partial t}\int_{\Omega}d^3\pmb{r}\pmb{j}(\pmb{r},t)=-\int_\Omega d^3\pmb{r}n(\pmb{r},t)\nabla\upsilon(\pmb{r},t)~,
\end{equation}
where $n(\pmb{r},t)$ is the time-dependent electron density of the material driven by laser field. External potential $\upsilon(\pmb{r},t)$ corresponds to both the electron-nuclei potential and the applied laser field. The harmonic spectrum [$I(\omega)$] is obtained by applying a discrete Fourier-transform ($\mathcal{FT}$) to $\pmb{J}(t)$, given as

\begin{equation}\label{hhg}
    I(\omega) = \bigg|\mathcal{FT}\Big(\pmb{J}(t)\Big)\bigg|^2~.
\end{equation}

\noindent

Time dependent dipole currents can capture the first signature of the anisotropic nonlinear response. With respect to the crystal symmetry, we analyze the polarization components of the harmonic radiation that are perpendicular and parallel to the linearly polarized fundamental field. For any particular laser polarization direction (along AC or ZZ direction), prominent electronic currents (using Equation \ref{electronic_current}) and corresponding harmonic radiation are obtained for both directions parallel and perpendicular to the laser polarization direction, due to the anisotropy in phosphorene crystal (elaborated in the next paragraph). The resulting electronic currents for pristine phosphorene are shown in Figure \ref{fig2}(c) and Figure \ref{fig2}(d).
In Figure \ref{fig2}(c), the electronic current parallel to the electric field direction (along \textit{x}) is much higher than the perpendicular contribution (along \textit{y}). Similarly, in Figure \ref{fig2}(d), when the laser polarization is along ZZ direction, the parallel current component (along \textit{y}) is much greater than the perpendicular current component (along \textit{x}). From Figure \ref{fig2}(c,d), it is evident that the parallel electronic current component is nearly three orders of magnitude higher than the perpendicular component. The laser polarization direction with respect to the crystal orientation is shown schematically in Figure \ref{fig2}(e,f). As shown, in AC configuration, the laser is polarized along AC direction, and likewise for ZZ. Each of these configurations consists of two different scenarios in which, w.r.t. the laser polarization direction, the parallel ($\parallel$) and perpendicular ($\perp$) polarization components of the harmonic radiation are obtained. When the laser is polarized along ZZ direction, for $\varepsilon=-10\%$, due to gross change in the crystal structure (as shown in Table \ref{geometry_parameters}) and consequently the electronic properties, the two current components are noted to have similar/comparable amplitude (see supplementary Figure \ref{Si2} ). 



HHG signal in both parallel and perpendicular directions w.r.t. laser polarization in phosphorene stems from its intrinsic anisotropy, which manifests inversion and mirror reflection symmetries only along the ZZ direction, as demonstrated in Figure \ref{fig2}(e,f). Due to this inversion symmetry, only odd harmonics are obtained parallel to the laser polarization direction \cite{Liu2016,Jia2020}. This symmetry conditions and symmetry-forbidden selection rules result in electronic currents along both \textit{x} and \textit{y}, when incident laser polarization is along AC; however the perpendicular current component is 3 orders of magnitude less than the parallel current component along \textit{x} (Figure \ref{fig2}c). 

\subsection{Mechanism of HHG}

\begin{figure}[t]
    \centering
    \includegraphics[width=.95\textwidth]{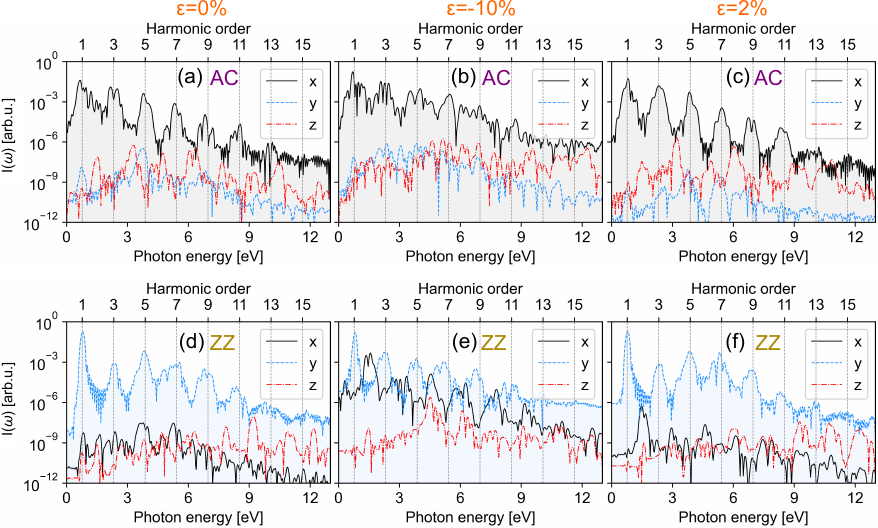}
    \caption{HHG spectra of pristine phosphorene (a,d), $-10\%$ strained (b,e), and $2\%$ strained (c,f) phosphorene systems  driven by a laser polarized along AC (a,b,c) and along ZZ (d,e,f).}
    \label{fig3}
\end{figure}

HHG spectra for pristine (Figure \ref{fig3}(a,d)), -10\% (Figure \ref{fig3}(b,e)), and 2\% (Figure \ref{fig3}(c,f)) strained phosphorene systems driven by laser polarized along AC (Figure \ref{fig3}(a-c)) and ZZ (Figure \ref{fig3}(d-f)) direction is investigated. 
Anisotropic HHG response is clearly visible by the distinctive harmonic radiation obtained for AC and ZZ cases in all the panels. In all the cases except, -10\% ZZ (as shown in Figure \ref{Si2}), the parallel component of the harmonics is much higher than the perpendicular component.

\begin{figure}
    \centering
    \includegraphics[width=.9\textwidth]{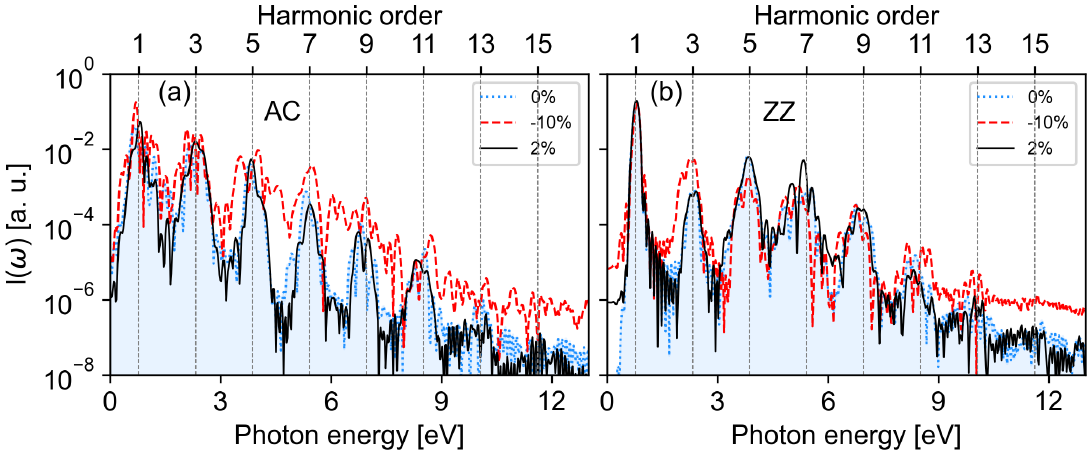}
    \caption{HHG spectra of pristine (dashed line), $-10\%$ (dash dotted line) and $2\%$ (dotted line) strained phosphorene systems driven by the laser polarized along AC (a) and ZZ (b) direction.}
    \label{fig4}
\end{figure}

Different features in HHG spectrum resolved in parallel and perpendicular directions for AC and ZZ configurations [in Figure \ref{fig4}(a,b)] originates from the intrinsic anisotropy of the phosphorene crystal.  For the parallel harmonic components, i.e., x in AC case [black lines in Figure \ref{fig3}(a-c)] and y in ZZ case (blue lines in Figure \ref{fig3}(d-f)], for both pristine and strained structures, the fundamental peak appears at $\omega$ ($=0.77 ~eV$), followed by odd integer multiples of the driving laser frequency [see Figure \ref{fig4}(a,b)], suggesting that  the inversion symmetry of the system is protected even by applying biaxial compressive and tensile strains. In each case, the HHG peak intensity is found to monotonically decrease, and is distinctly visible until H13. 
In both AC and ZZ cases, compressive strain ($\varepsilon=-10\%$) leads to significant enhancement in the harmonic intensity compared to the pristine case. Additionally, in AC case, for $\varepsilon=-10\%$, HHG peaks become stronger and spectral width broadens, as can be seen in Figure \ref{fig4}(a), while for ZZ there is no systematic enhancement. 
Interestingly, in -10\% case, from the finite electronic population at the Fermi level [as evident along $\overline{\Gamma X}$ and $\overline{\Gamma Y}$ directions in Figure \ref{fig1}(c,d)], some states are always available for field-induced excitations. Thus, even when the field is along ZZ, prominent current and HHG signal can be obtained along perpendicular direction.
As found in Figure \ref{Si1}(b), along $\overline{\Gamma X}$ (i.e. AC direction), electronic states at and around the Fermi level ($E_F$) contribute significantly to field induced transition, thereby resulting in higher strain-induced HHG enhancement. For example, in the AC configuration for $-10\%$ case, the high harmonic spectrum [from H5 in Figure \ref{fig4}(a)] clearly shows the harmonic intensity enhancement by nearly an order of magnitude. In particular for -10\% case, due to significant structural changes (see Table \ref{geometry_parameters}), the effective electron-nuclei potential is prominently modified (in Equation \ref{electronic_current}), resulting in distinctive new features in the electronic band dispersion, thereby affecting the microscopic electronic current, and eventually the HHG signal.
The lower the states are in the conduction band (CB), the easier it is for the free carriers to get excited by the laser field from the VBM to these lowered CB states, which eventually leads to an enhanced HHG \cite{Qin2018}. Furthermore, due to the applied strain, the conduction bands and the valence bands shift downwards. This downward shift of bands is more prominent in the case of $\varepsilon=-10\%$ compared to $\varepsilon=2\%$. However, in $\varepsilon=-10\%$ case, the available electronic states in the downward shifted VBM along $\overline{\Gamma Y}$ are low compared to that in $\varepsilon=2\%$ case. As a consequence, the harmonic yield for the ZZ configuration is expected to be higher in the case of $\varepsilon=2\%$ compared with the rest of the cases, which is investigated and discussed later. In order to clearly understand these results, we show the computed number of electrons excited to the conduction band from the valence band during the laser pulse in Figure \ref{fig5} corresponding to Figure \ref{fig4}.

\subsection{Interband and intraband contributions}
It is well known that in a semiconductor, recombination of an electron with an existing hole in the valence band results in the emission of a photon with energy exceeding the band gap energy. Consequently, it becomes evident that harmonics below the band gap do not stem from the interband contribution. Hence, below bandgap harmonics are only from intraband current. However, above band gap harmonics can originate from both interband and intraband currents. In case of phosphorene, the calculated band gap is $0.9~ eV$, and induced photon energy is $0.77~ eV$. Thus, in this scenario, recombination of electrons and holes in different bands becomes possible, and both interband and intraband processes are allowed to jointly contribute to the generation of harmonics above the band gap.
To estimate the extent of interband transitions in our systems of study, we calculated the temporal evolution of total number of excited electrons from valence band to conduction band by projecting the time-evolved wave functions ($\ket{\psi_{i,\textbf{k}}(t)}$) on to the basis of ground state wave functions ($\ket{\psi_{j,\textbf{k}}^{GS}}$), given as \cite{TancogneDejean2017NC}

\begin{equation}\label{exci_elec}
    N_{ex}(t) = N_e - \frac{1}{N_\textbf{k}}\sum_{i,j}^{occ}\sum_\textbf{k}^{BZ}|\langle\psi_{i,\textbf{k}}(t)|\psi_{j,\textbf{k}}^{GS}\rangle|^2~,
\end{equation}

\noindent where $N_e$, $N_\textbf{k}$, $\psi_{i,\textbf{k}}$, and $\psi_{j,\textbf{k}}$ are the total number of electrons in the system, the total number of \textbf{k}-points used to sample the BZ, Kohn-Sham state at the \textit{i}-th band at the \textbf{k}-point \textbf{k}, and the ground Kohn-Sham state at the \textit{j}-th band at the \textbf{k}-point \textbf{k}, respectively. The sum over the band indices $i$ and $j$ run over all occupied states and the summation runs over the entire first Brillouin zone.

\begin{figure}[H]
    \centering
    \includegraphics[width=.9\textwidth]{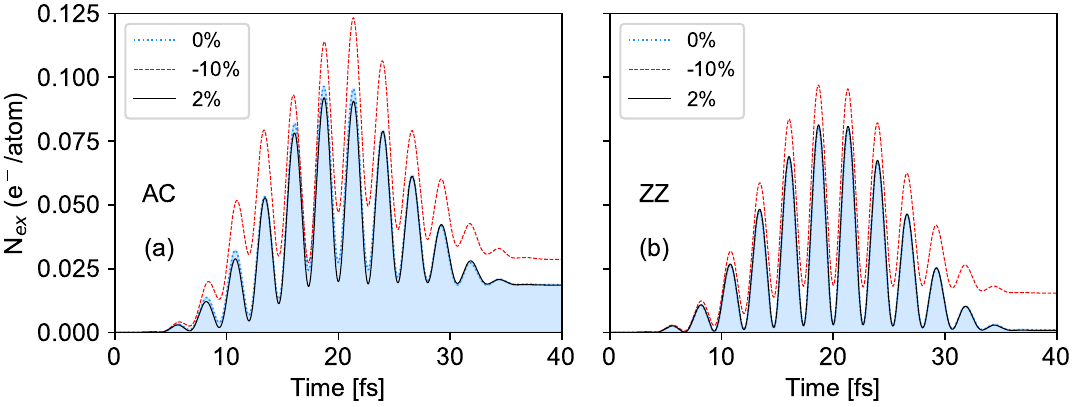}
    \caption{Total number of excited electrons from valence bands to the conduction bands during the laser pulse, which is polarized along AC (a) and along ZZ (b) in pristine (in blue), $-10\%$ (in red), and $2\%$ (in black) strained phosphorene. 
    }
    \label{fig5}
\end{figure}

Thus, $N_{ex}(t)$ indicates the amount of interband current. Temporal evolution of $N_{ex}$ displays same sub-cycle periodicity as the induced laser field (Figure \ref{fig5}(a) for AC , Figure \ref{fig5}(b) for ZZ). 
Figure \ref{fig5} shows that the number of excited electrons is larger in the case of AC-polarized laser, which can hence lead to larger electronic current and higher harmonic yield along AC (consistent with Figure \ref{fig4}).

It is important to note that the anisotropic electronic band dispersion shows flatter bands along $\overline{\Gamma Y}$ (ZZ), in comparison to $\overline{\Gamma X}$ (AC direction). This implies lower effective mass and higher carrier mobility in AC direction, in comparison to ZZ. Consequently interband current is prominently larger in AC direction, as demonstrated by corresponding $N_{ex}$ (compare both panels in Figure \ref{fig5}). Consequently, significant difference in the charge excitation along different directions is expected. For -10\% strain in  AC case, the laser induced excited electrons reaches a peak value of 0.12 $e^-$/atom (0.48 $e^{-}$/unit cell) at $\sim21 ~fs$ and then oscillates at around 0.035 $e^-$/atom (0.14 $e^-$/unit cell) [Figure \ref{fig5}(a)]. Similar value of $e^-$/atom is obtained for 0\% and 2\% strain cases, with 0\% slightly higher than 2\%. For ZZ configuration, the maximum excited $e^-$/atom of 0.10 is attained for -10\% strain case and then oscillates at 0.02 by the end of the laser pulse. 
However, for 0\% and 2\% strain cases, the excited electrons (slightly higher in 2\% case compared to 0\% case) returns to almost zero after each half cycle in the ZZ configuration, while for AC configuration for all cases, a fraction of excited electrons remain effectively excited to conduction bands at the end of the laser pulse ($\sim 40 ~fs$). 
These strain-induced number of excited electrons are well in agreement with the corresponding harmonic spectra shown in Figure \ref{fig4}. 
 Only for -10\% strain case, for both laser polarization along AC and ZZ direction, during each half cycle, more electrons are excited, however fraction of excited electrons do not return to ground state. This indicates that carriers have been driven to higher energy levels (interband) and then remained there even after the removal of the laser pulse. These excited electrons must occupy broader areas of BZ, i.e., larger momentum space, thereby resulting in higher intraband current under compressive strain, which contribute to the harmonic generation collectively with interband current.


\begin{figure}[t]
    \centering
    \includegraphics[width=.7\textwidth]{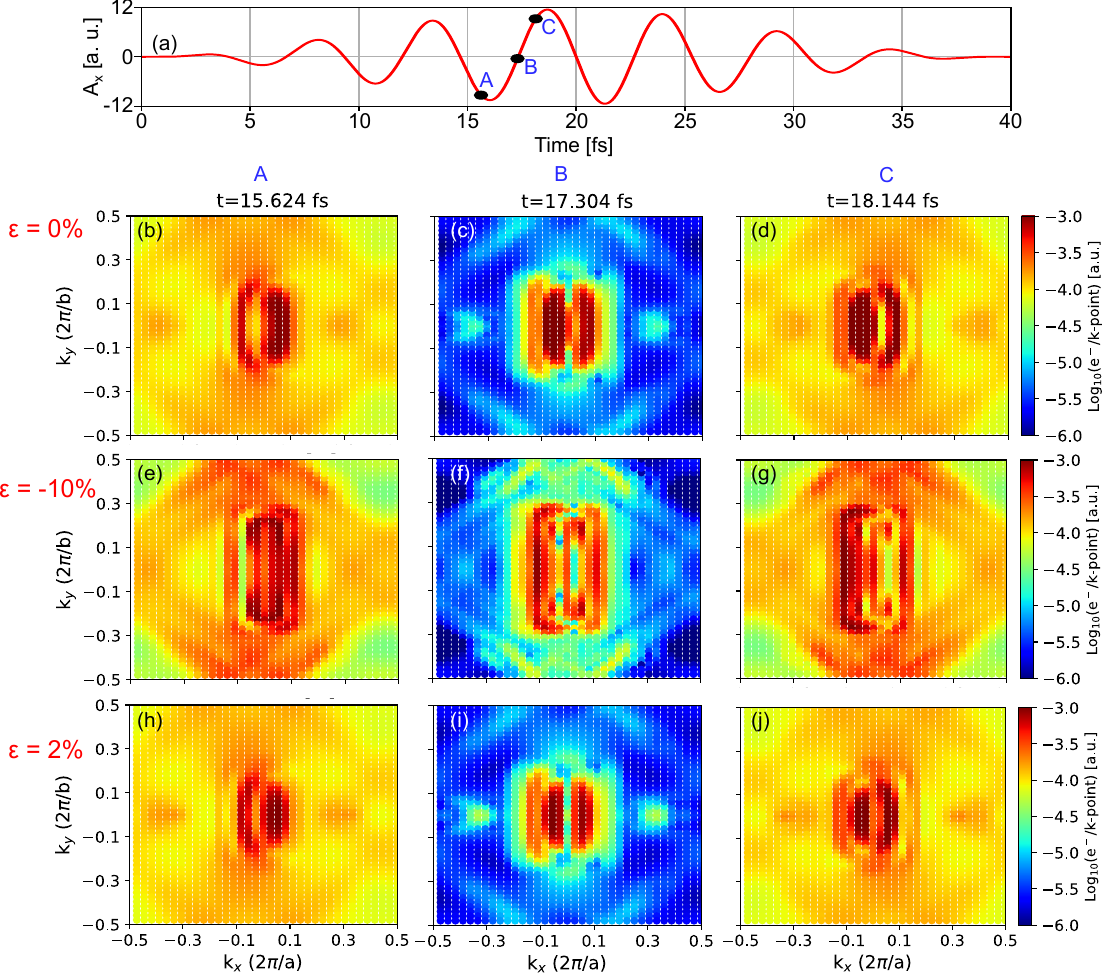}
    \caption{Time-dependent electron dynamics in strain-free and strained phosphorene systems. (a) Vector potential of the driving laser pulse polarized along AC (\emph{x}-direction). Snap shots of the sub-cycle dynamics of momentum-resolved excited electrons near the peak and minimum of the vector potential (marked as A, B, C in (a)) for pristine phosphorene (b,c,d), -10\% (e,f,g), and 2\% (h,i,j) strained phosphorene. The oscillation features of the excited electrons in the BZ in panels (b-d) for pristine, (e-g) for -10\%, and (h-j) for +2\% show evidence of intraband process significantly contributes to HHG.}
    \label{fig6}
\end{figure}
Furthermore, to validate our interpretation and to obtain laser-driven subcycle electron dynamics, we calculate the momentum-resolved excited electrons distribution $N_{ex}(\textbf{k};t)$ in the following way \cite{TancogneDejean2017NC}:

\begin{equation}\label{mom-res}
    N_{ex}(\textbf{k};t) = \frac{1}{N_\textbf{k}}\bigg( N_e - \sum_{i,j}^{occ}|\langle\psi_{i,\textbf{k}}(t)|\psi_{j,\textbf{k}}^{GS}\rangle|^2\bigg)~.
\end{equation}

We compare the momentum-resolved sub-cycle dynamics of excited electrons at three time instances marked as A, B, C in Figure \ref{fig6}(a), which are near the maximum and minimum of the vector potential, which is polarized along AC direction for 0\% (top), $-10\%$ (middle), and $2\%$ (bottom) strained phosphorene systems shown in Figure \ref{fig6}(b-j). It is interesting to observe that the excited electron density is concentrated at the $\Gamma-$point for all the cases. For both 0\% and 2\% strained phosphorene, both VBM and CBM are at $\Gamma-$point, hence electron transition probability is maximum at $\Gamma$ (due to their similar band structure). On the other hand, for -10\% strained system, VBM and CBM overlap near $\Gamma-$point (see $\overline{\Gamma X}$ band dispersion in Figure \ref{fig1}c). This results in the higher concentration of excited electron density at and around the $\Gamma-$point in this particular case.  
At time instant denoted by B in Figure \ref{fig6}(a), i.e. when the system has the lowest excitation, electrons wander to broader areas in BZ under compressive strain [see Figure \ref{fig6}(f)]. Hence, the number of excited electrons at each momentum ($\textbf{k}_x,\textbf{k}_y$) increases, which confirms that intraband and interband processes cooperatively enhances the harmonic yield for -10\% case. At the time instants indicated by A and C in Figure \ref{fig6}(a), a large part of BZ has larger amount of excited electrons, and it is clear that the excited electrons arrange themselves from left(right) to right(left) boundaries in the BZ. These sub-cycle oscillations of excited electrons driven by the laser field along with excited electrons traversing to larger parts of BZ shows that intraband and interband processes collectively contributes to HHG and enhanced harmonic yield compared to pristine and strained phosphorene cases, which is highlighted later.

\begin{figure}
    \centering
    \includegraphics[width=.95\textwidth]{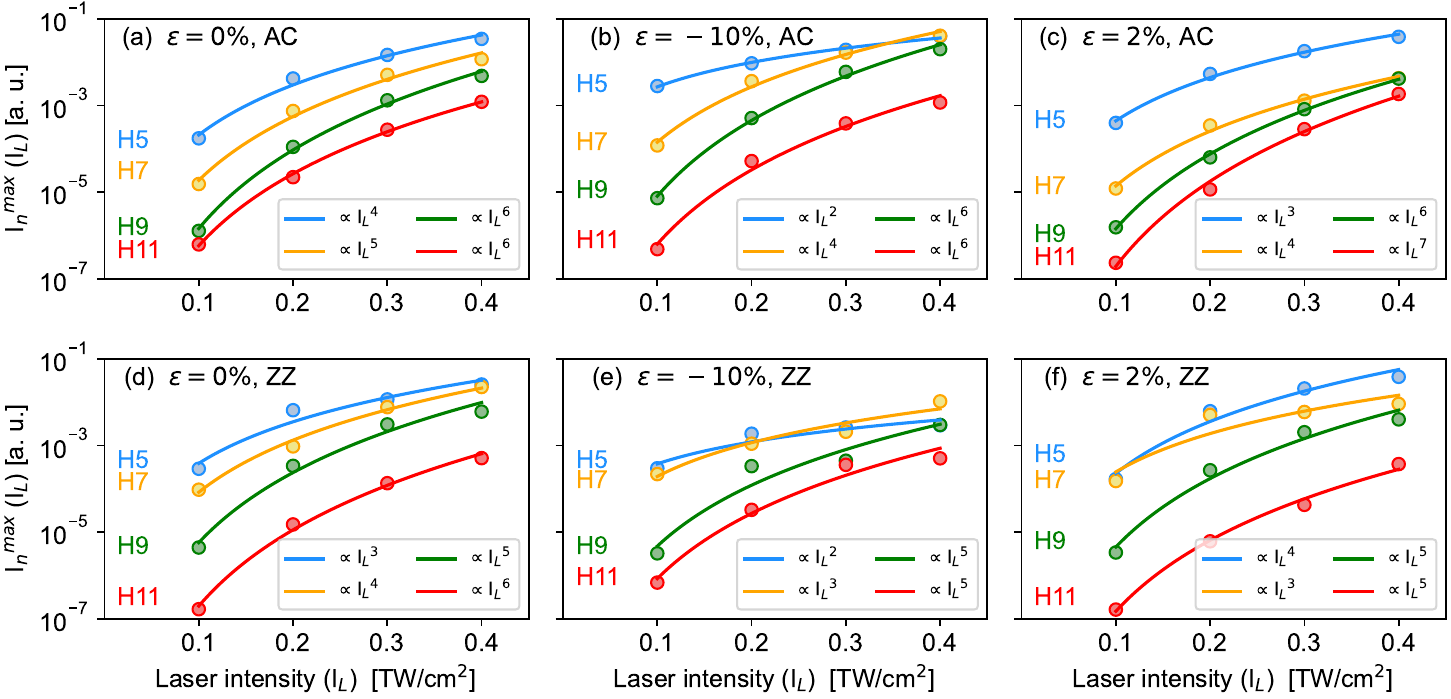}
    \caption{Calculated peak harmonic intensity as a function of peak laser intensity $I_L$ for 0\% strain (a,d), -10\% (b,e), and 2\% strain (c,f) for representative harmonics n $= 5$ (blue circles), 7 (orange circles), 9 (green circles), and 11 (red circles) obtained for laser polarized along AC (a-c) and along ZZ (d-f). Solid lines are obtained by fitting data to power law, yielding exponents showing the non-perturbative scaling of the harmonic process. The resulting exponents are mentioned in the legends of individual panel. 
    }
    \label{fig7}
\end{figure}

To reveal the nature of the HHG spectrum, we calculated the peak harmonic intensity ($I_n^{max}$) of representative harmonics ($n$) as a function of peak laser intensity ($I_L$). Figure \ref{fig7} illustrates the scaling behavior of harmonic peaks $n=$ 5, 7, 9 and 11 as a function of $I_L$, ranging from $0.1 - 0.4~ TWcm^{-2}$. In all the configurations shown in Figure \ref{fig7}, for $n=5, ~7$, fit to the power law $I^p$ resulted in \emph{p} with a value that ranges between $2 - 5$. Whereas, for the high-order harmonics, for instance H9 and H11, \emph{p} ranges between $5 - 7$. By fitting the data to the power-law, the non-perturbative nature of the high-order harmonics is clearly evident in the scaling of individual harmonic peaks with $I_L$. Note that in the perturbative regime, the radiation is limited to frequencies that are merely a few times higher than the frequency of the driving field. Furthermore, the $n^{th}$ harmonic would scale proportionally with $I^n_L$, where $I_L$ is the peak intensity of the laser and the harmonic yield would reduce exponentially \cite{Goulielmakis2022,Liu2016}. 

\begin{figure}[H]
    \centering
    \includegraphics[width=.95\textwidth]{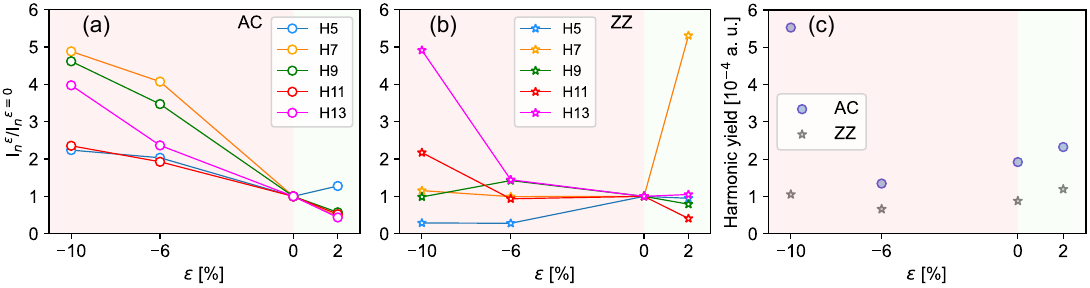}
    \caption{The relative change in harmonic intensity as a function of applied strain for representative harmonics for (a) AC case and (b) ZZ case. (c) Harmonic yield (integrated from H2 to H15) as a function of applied strain for both AC and ZZ cases.}
    \label{fig8}
\end{figure}


For each harmonic peak, the relative change in the harmonic intensity and total harmonic yield are compared for AC and ZZ cases, as shown in Figure \ref{fig8}. The relative change in the harmonic intensity is defined as $I^{\varepsilon}_n/I^{\varepsilon=0}_n$, where $I^{\varepsilon}_n$ and $I^{\varepsilon=0}_n$ stands for $n^{th}$ order harmonic peak intensity with strain and without strain, respectively. Consistent with HHG signals (in Figure \ref{fig4}), for laser polarized along AC (Figure \ref{fig8}a), the compressive strain of -10\% enhances HHG intensities by $100-400\%$ for all harmonic orders, H7 being most prominently affected followed by H9, H13, H11 [see Figure \ref{fig8}(a)]. With tensile strain, however, the relative harmonic intensity of all the  harmonics, except for $5^{th}$ order, decreased, as seen in Figure \ref{fig8}(a). 
For laser polarization along ZZ, each harmonic responded differently with respect to the applied strain (Figure \ref{fig8}(b)), reflecting the lack of systematic variation of HHG intensity. 
Different HHG peak intensities and harmonic yield for AC and ZZ cases is consistent with anisotropic band dispersion of phosphorene and distinctive features.
Harmonic yield (integrated from 2nd order to 15th order) as a function of applied strain (Figure \ref{fig8}(c)) shows almost 3 times enhancement in total yield when $-10\%$ strain is applied for AC case. This demonstrates the possibility of tailoring the HHG yield using both appropriate strain and laser electric field direction. 

\section{Conclusions}\label{conclusions}
In summary, using state-of-the-art TDDFT simulations, we have demonstrated the effects of ultrashort laser induced non-linear response and yield of non-perturbative HHG from a monolayer black phosphorene, extending to the 13th order.
In particular the role of biaxial compressive and tensile strain on the high harmonic generation from monolayer phosphorene is investigated. Our predictions show that tuning the crystal structure of phosphorene via biaxial strain provides an efficient tool to tailor its electronic structure and HHG in terms of harmonic intensity and spectral width. Due to the intrinsic in-plane anisotropy of phosphorene, the harmonic yield for the AC configuration is higher compared to that obtained for the ZZ configuration. The laser pulse parameters as chosen in our study are in Mid-IR regime, 
that are suitable to probe HHG from solid-state band-gap materials, and available in advanced laser facilities like 
ELI ALPS facility \cite{Khn2017,Charalambidis2017}. Our analysis shows that, the harmonic yield can be enhanced by a factor of 3 by applying $-10\%$ compressive strain. The origin of this unique response can be attributed to the electronic band structure engineering through external compressive strain, that results in prominent increase in the number of  excited electrons, which eventually leads to an enhanced HHG. Our results and insights promotes the possibility to tune the HHG process by band engineering through the external application of strain and also opens the door to experimentally investigate strain-induced strong-field nonlinear response of this promising 2D nanostructure.

In light of the growing enthusiasm for developing compact table-top sources of coherent extreme ultraviolet XUV radiation, which hold promise for practical applications across various domains of science and technology, and given the ever-expanding spectrum of potential applications for 2D materials, our study showcases the promise of phosphorene and analogous 2D nanostructures. With easily tunable band structures feasible through strain engineering, these 2D materials exhibit significant potential for designing and optimizing innovative devices in the field of attoscience, as well as for the creation of stable table-top HHG sources. 


\section*{Declaration of Competing Interest}
The authors confirm that they have no competing financial interests or personal relationships that could have influenced the findings presented in this paper.

\section*{Acknowledgements}
The ELI ALPS project (GINOP-2.3.6-15-2015-00001) is supported by the European Union and it is co-financed by the European Regional Development Fund. This research has been supported by the IMPULSE project which receives funding from the European Union Framework Programme for Research and Innovation Horizon 2020 under grant agreement No 871161. SK, MUK and SM also acknowledges project No. 2019-2.1.13-T\'ET-IN-2020-00059, which has been implemented with support provided by the National Research, Development and Innovation Fund of Hungary, and financed under the 2019-2.1.13-T\'ET-IN funding scheme. MUK and SM would like to thank the ELI ALPS IT department and the HPC administration for their support in providing computational resources. This research is supported by the \'{U}NKP-23-4 - New National Excellence Program of the Ministry for Culture and Innovation from the Source of the National Research, Development and Innovation Fund.
\clearpage

\printbibliography
\clearpage

\renewcommand{\appendixname}{Supplementary Information}

\appendix
\renewcommand{\thesection}{Supplementary Information \arabic{section}}    
\renewcommand{\thefigure}{SI~\arabic{figure}}
\setcounter{figure}{0}
\renewcommand{\thetable}{SI~\arabic{table}}
\setcounter{table}{0}




\section{Projected bands and partial density of states}
\begin{figure}[H]
    \centering
    \includegraphics[width=.9\textwidth]{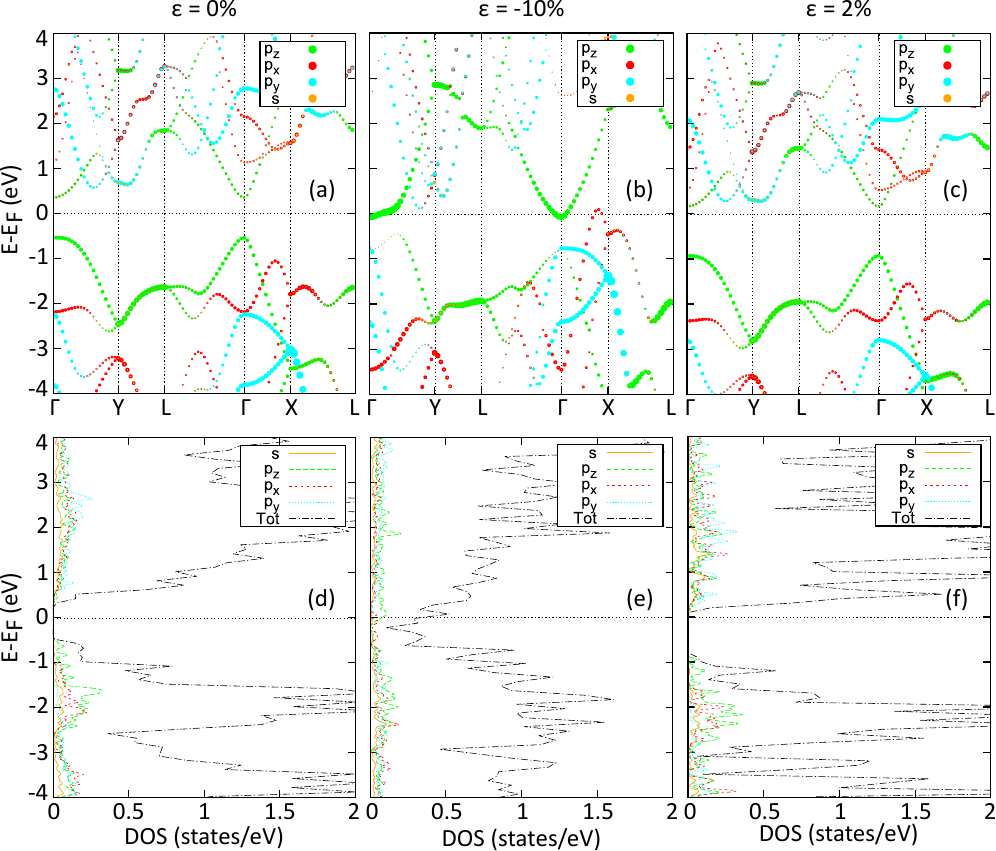}
    \caption{(a-c) Projected band structure and (d-f) partial density of states of 0\%, -10\%, and 2\% strained phosphorene systems.}
    \label{Si1}
\end{figure}

\section{Induced electronic currents in -10\% strained phosphorene.}
\begin{figure}[H]
    \centering
    \includegraphics[width=.9\textwidth]{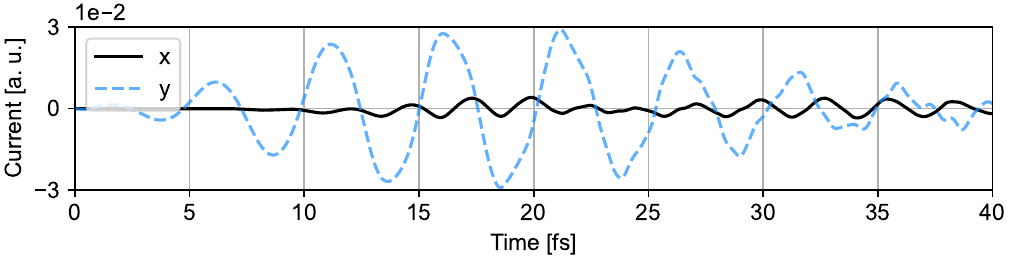}
    \caption{Induced electronic currents in $-10\%$ strained phosphorene that are perpendicular (\textit{x}) and parallel (\textit{y}) to the incident laser polarization along ZZ direction.}
    \label{Si2}
\end{figure}

\section{Total excited electrons per atom at the end of the laser pulse}
\begin{figure}[H]
    \centering
    \includegraphics[width=.65\textwidth]{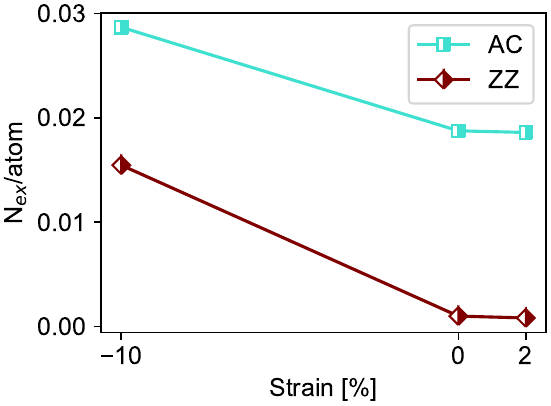}
    \caption{Dependence of biaxial strain on the total number of excited electrons at the end of the laser pulse.}
    \label{Si3}
\end{figure}


\end{sloppypar}
\end{document}